\documentstyle[12pt,epsf]{article}
\newcommand{\beq}{\begin{equation}}
\newcommand{\eeq}{\end{equation}}
\newcommand{\beqa}{\begin{eqnarray}}
\newcommand{\eeqa}{\end{eqnarray}}

\begin{document}

\pagestyle{empty}

\hfill TK 95 28

\smallskip

\begin{center}

{\large { \bf NUCLEON FORM FACTORS FROM DISPERSION THEORY}}

\vspace{1.cm}

Ulf--G. Mei{\ss}ner\\
{\it Universit\"at Bonn, ITKP, Nussallee
14-16, D--53115 Bonn, Germany}

\vspace{0.4cm}

\end{center}

\baselineskip 10pt

\noindent I review the results of a recent dispersion--theoretical analysis
of the nucleon electromagnetic form factors and comment on the strangeness
form factors. The need for a better data basis at low, intermediate and large
momentum transfer is stressed.

\vspace{0.5cm}

\baselineskip 14pt

\noindent {\bf 1 $\quad$ WHY DISPERSION RELATIONS ?}

\vspace{0.3cm}

\noindent The structure of the nucleon (denoted by '$N$')
as probed with virtual photons
is parametrized in terms of four form factors,
\beq
<N(p')\, | \, {\cal J}_\mu \,  | \, N(p)>
= e \,  \bar{u}(p') \, \biggl\{  \gamma_\mu F_1^{p,n} (t)
+ \frac{i \sigma_{\mu \nu} k^\nu}{2 m_N} F_2^{p,n} (t) \biggr\} \,  u(p) \,\, ,
\eeq
with $t = k_\mu k^\mu = (p'-p)^2$ the invariant momentum
transfer squared, ${\cal J}_\mu$
the em current related to the photon field and $m_N$ the nucleon mass.
In electron scattering, $t < 0$ and it is thus convenient
to define the positive
quantity $Q^2 = -t > 0$. $F_1$ and $F_2$ are called the Pauli and the Dirac
form factor (ff), respectively. There exists already a large body of data
for the proton and also for the neutron. In the latter case, one has to perfrom
some model--dependent extractions to go from the deuteron to the neutron.
More accurate data are soon coming (ELSA, MAMI, CEBAF, $\dots$). It is
thus mandatory to have a method which allows to analyse all these data in a
mostly model--independent fashion. That's were dispersion theory comes into
play. Although
not proven strictly (but shown to hold in all orders in perturbation
theory), one writes down an unsubtracted dispersion relation for $F(t)$ (which
is a generic symbol for any one of the four ff's),
\beq
F(t) = \frac{1}{\pi} \int_{t_0}^\infty \, dt' \, \frac{{\rm Im} \, F(t)}{t'-t}
\, \, , \eeq
with $t_0$ the two (three) pion threshold for the isovector (isoscalar) ffs.
Im~$F(t)$ is called the spectral function. These spectral functions are
the natural meeting ground for theory and experiment, like e.g. the partial
wave amplitudes in $\pi N$ scattering.
If the data were to be infinitely precise, the
continuation from negative $t$ (data) to positive $t$ (spectral functions)
in the complex--$t$ plane would lead to a unique result for the
spectral functions. Since that is not the case, one has to make some extra
assumption guided by physics to overcome the ensuing instability as will be
discussed below. Let me first enumerate the various constraints one has for the
spectral functions.

\vspace{0.5cm}

\noindent {\bf 2 $\quad$ CONSTRAINING THE SPECTRAL FUNCTIONS}

\vspace{0.3cm}

\noindent In general, the spectral functions can be thought
of as a superposition
of vector meson poles and some continua, related to n-particle thresholds, like
e.g. $2\pi$, $3\pi$, $K \bar{K}$ and so on. For example, in the Vector Meson
Dominance picture one simply retains a set of poles. However, there are some
powerful constraints which these spectral functions have to obey:

\begin{itemize}

\item[$\bullet$] {\it Unitarity}: As pointed out by Frazer and Fulco [1] long
time ago, extended unitarity leads to a drastic enhancement of the isovector
spectral functions on the left--wing of the $\rho$ resonance. Leaving out this
contribution from the two--pion cut leads to a gross underestimation of the
isovector charge and magnetic radii. This very fundamental constraint is very
often overlooked. It is believed that in the three pion (isoscalar) channels
no such enhancement exists.

\item[$\bullet$] {\it pQCD}: Perturbative QCD (pQCD) tells us how the nucleons
ffs behave at very large momentum transfer based on dimensional counting
arguments supplemented with the leading logs due to QCD [2]. This leads to a
set
of superconvergence relations for Im~$F_1 (t)$, Im~$F_2 (t)$ and
Im~$t \, F_2 (t)$, which have to be imposed ($F_2$ is suppressed by one more
power in $t$ than $F_1$ due to the spin--flip).

\item[$\bullet$] {\it The neutron radius}: Over the last years, the charge
radius of the neutron has been
determined very accurately by  measuring the neutron--atom
scattering length,
i.e. $F_1^n$ at $Q^2 = 0$. This value, which we take from the recent paper
[3], has to be imposed as a further constraint.

\item[$\bullet$] {\it Stability}: The isovector spectral functions are
completely fixed from $t=(4 \ldots 50) \,M_\pi^2$ due to unitarity.
At large $t$, pQCD determines the behaviour of all isovector/isoscalar
spectral functions. In additon, we have a few more isovector and
isoscalar poles. Loosely spoken, their number is minimized by the requirement
that the data can be well fitted (for details, see Refs.[4,5]).

\end{itemize}

\vspace{0.5cm}

\noindent {\bf 3 $\quad$ RESULTS AND DISCUSSION}

\vspace{0.3cm}

\noindent The spectral functions fulfilling all the abovementioned requirements
consist of a hadronic part (the $2\pi$ continuum plus three additional
isovector and isoscalar poles) and the quark contribution leading to the pQCD
behaviour (parametrized by a simple log function which depends on a parameter
$\Lambda^2$ that can be considered a measure of
separating these two regimes) [4]. At this point, we have 8 free parameters.
In contrast to a previous dispersive analysis [5],
we are able to identify all three
isoscalar and two of the isovector masses with physical ones. Only the third
isovector mass is so tightly fixed by the constraints that it can not be chosen
freely. This leaves us with effectively three fit parameters. The best fit
to the nucleon form factors is shown in Fig.1 (to be precise, we show the
ffs normalized to the dipole fit, in case of $G_E^n$ we normalize to the
Platchkov data [6] for the Paris potential).
 From these, we deduce the following nucleon electric (E) and
magnetic (M) radii [4]:
\beq
r_E^p = 0.847~{\rm fm} \, , \, \,
r_M^p = 0.836~{\rm fm} \, , \, \,
r_M^n = 0.889~{\rm fm} \, , \, \,
\eeq
all with an uncertainty of about 1\%. These results are similar to the
ones found by H\"ohler et al. [5] with the exception of $r_M^n$ which
has increased by 5\% (due to the neglect of one superconvergence relation in
Ref.[5]). From the residua at the two lowest isovector poles, we can determine
the $\omega NN$ and $\phi NN$ coupling constants,
\beq
\frac{g_{\omega NN}^2}{4\pi} = 34.6 \pm 0.8 \, , \,\, \, \kappa_\omega
= -0.16 \pm 0.01 \, , \, \,
\frac{g_{\phi NN}^2}{4\pi} = 6.7 \pm 0.3 \, , \,\, \, \kappa_\omega
= -0.22 \pm 0.01 \, ,
\eeq
where $\kappa_V$ ($V = \omega, \phi$) is the tensor to vector coupling
strength ratio. These results are similar to the ones in Ref.[5].

\vspace{0.3cm}

\hskip 1.in
\epsfysize=5in
\epsffile{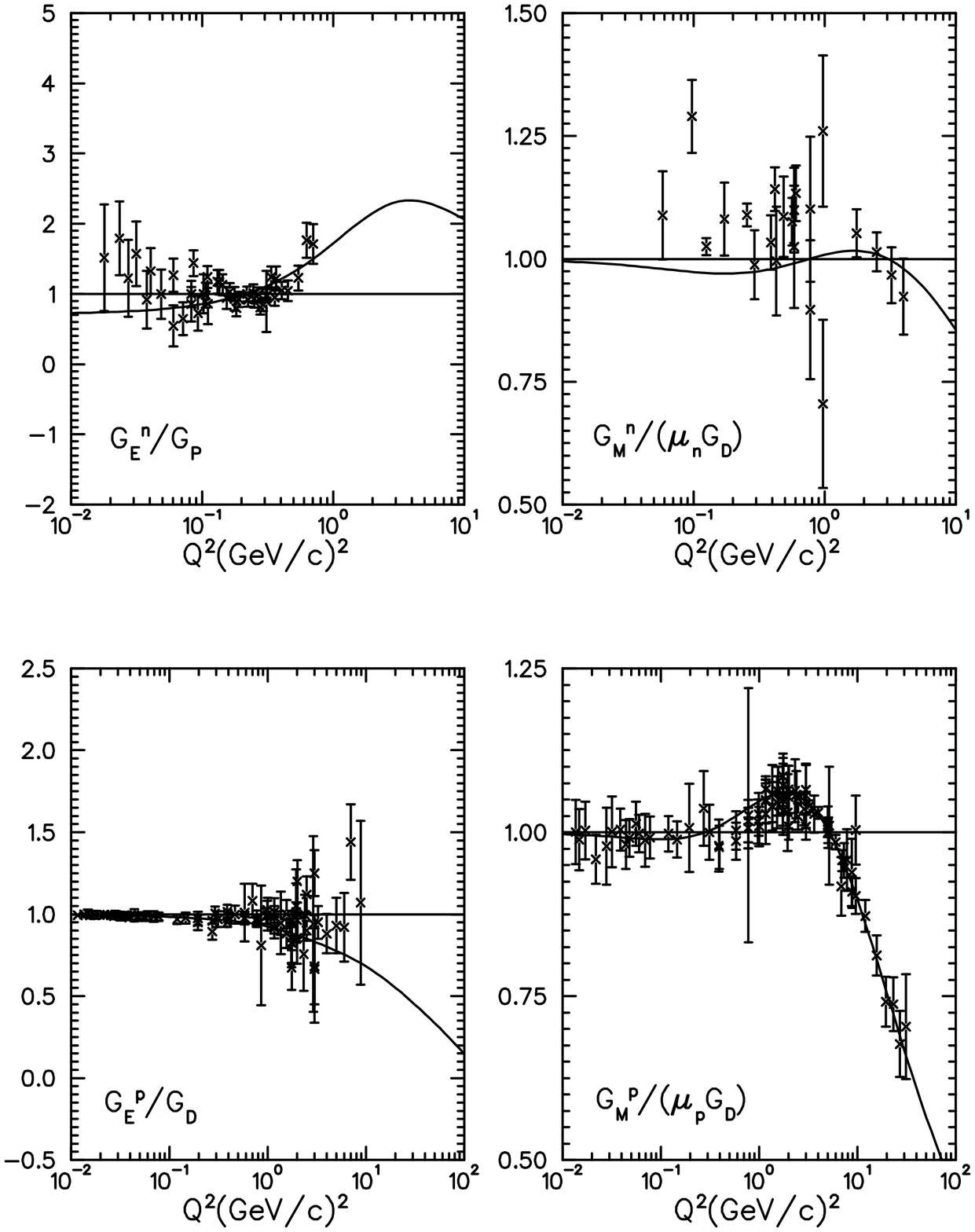}


\centerline{Fig.~1:\quad Best fit to the nucleon em form factors}


\medskip \bigskip

\noindent Of particular interest is the
onset of pQCD. Only for $G_M^p(t)$ data for
$Q^2 > 10$~GeV$^2$ exist. While these data are consistent with the pQCD
scaling $L^{-1}(Q^2) Q^4 G_M^p(Q^2) \to \, \,  $constant, where $L^{-1}(Q^2)$
accounts for the leading logs, they are not precise enough to rule out a
non--scaling behaviour, see the lower panel in Fig.2. Also shown in Fig.2
is the same quantity without the log corrections (upper panel).
 All data for the much discussed quantity $Q^2 F_2^p
(Q^2) / F_1^p (Q^2)$ are below $Q^2 = 10$~GeV$^2$ which in our approach is
still in the hadronic region since $\Lambda^2 \simeq 10$~GeV$^2$ for the
best fit.

\vspace{0.5cm}

\noindent {\bf 4 $\quad$ STRANGE FORM FACTORS}

\vspace{0.3cm}

\noindent Jaffe [7] has shown how one can get bounds on the strange vector
form factors in the nucleon from such dispersion theoretical results. The
main assumption of his approach is that these strange form factors have the
same large--$t$ behaviour as the non--strange isoscalar ones. If the fall--off
for the strange form factors is faster, the strange matrix elements will be
reduced. Using our best fit together with a better treatment of the symmetry
breaking in the vector nonet, it is
straightforward to  update Jaffe's analysis. We find for the
strange magnetic moment and the strangeness radius [8],
\beq
\mu_s = -0.24 \pm 0.03 \, \, {\rm n.m.} \, , \, \, \,
r_s^2 = 0.21 \pm 0.03  \, \, {\rm fm}^2 \, .
\eeq
Furthermore, the strange ff $F_2^s (t)$ follows a dipole with a cut--off mass
of 1.46~GeV, $F_2^s (t) = \mu_s /(1- t /2.41 \, {\rm GeV}^2)^2$. It is
important
to stress that these numbers should be considered as upper bounds.

\vspace{0.5cm}

\hskip 1.9in
\epsfysize=3in
\epsffile{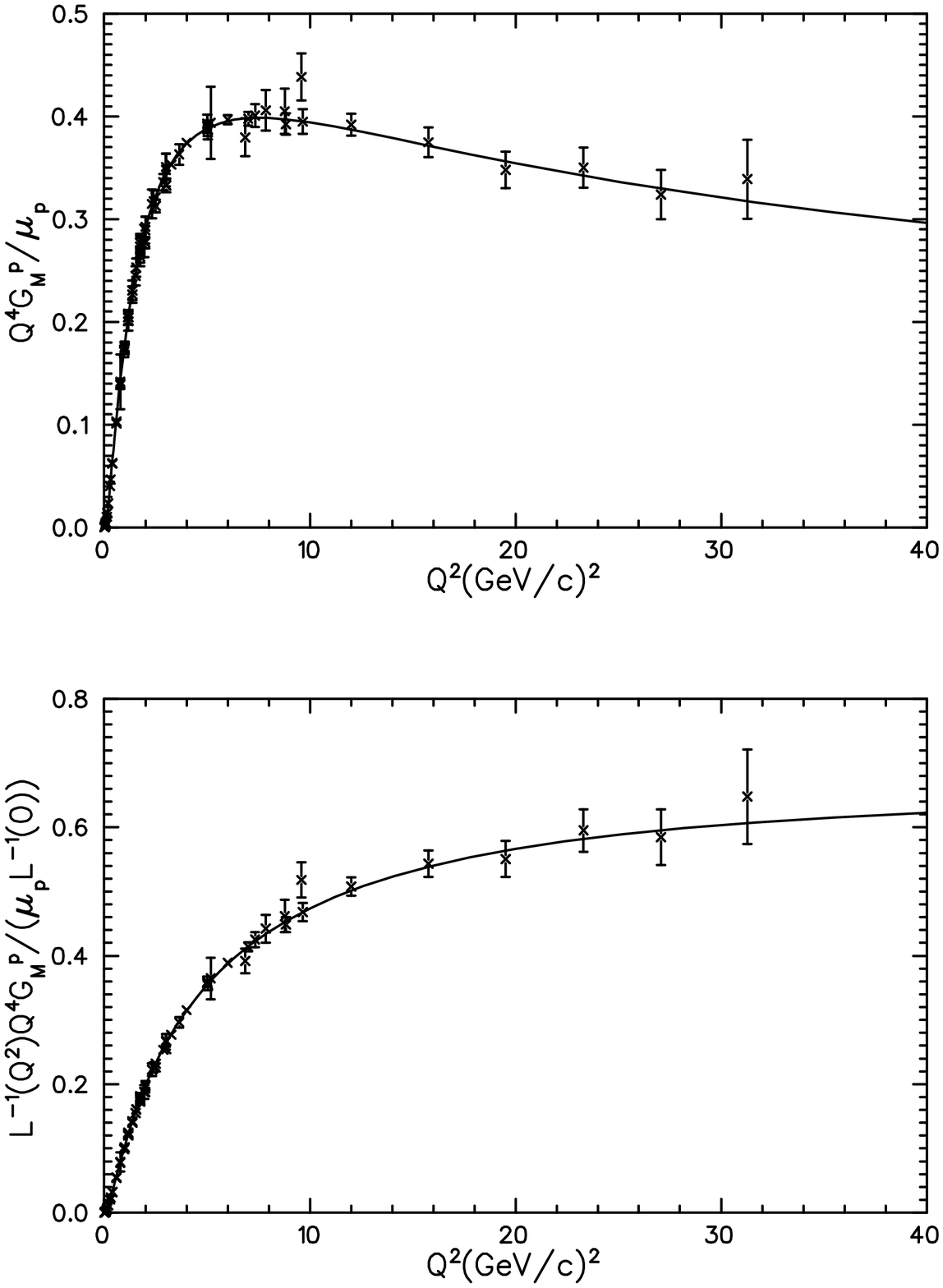}


\centerline{Fig.~2:\quad  pQCD scaling in $G_M^p(Q^2)$? Upper/lower panel:
Without/with leading logs.}


\medskip 

\vspace{0.4cm}

\noindent {\bf 5 $\quad$ SUMMARY}

\vspace{0.2cm}

\noindent The dispersion theoretical machinery has been updated to include
theoretical concepts like pQCD scaling and so on [4].
Now we need a more {\it accurate}
data basis at {\it low}, {\it intermediate} and
{\it large} momentum transfer to further
sharpen the extractions of the nucleon em radii, the $VNN$ coupling constants
and to  pin down the onset of perturbative QCD.

\vspace{0.4cm}

\noindent {\bf 6 $\quad$ ACKNOWLEDGEMENTS}

\vspace{0.2cm}

\noindent I am grateful to Dieter Drechsel, Hans--Werner Hammer and
Patrick Mergell for pleasant collaborations.

\vspace{0.4cm}

\noindent{\bf References}

\bigskip

\noindent 1. W.R. Frazer and F.J. Fulco, {\it Phys. Rev. Lett.}
{\bf 2}, 365 (1959).

\smallskip

\noindent 2. S.J. Brodsky and G. Farrar,
{\it Phys. Rev.} {\bf D11}, 1309 (1975).

\smallskip

\noindent 3. S. Kopecky et al., {\it Phys. Rev. Lett.}
{\bf 74}, 2427 (1995).

\smallskip

\noindent 4. P. Mergell, Ulf-G. Mei{\ss}ner and D. Drechsel,
{\it Nucl. Phys.\/} {\bf A} (1995) in print.

\smallskip

\noindent 5. G. H\"ohler et al., {\it Nucl. Phys.\/} {\bf B114}, 505 (1976).

\smallskip

\noindent 6. S. Platchkov et al., {\it Nucl. Phys.\/} {\bf A510}, 740 (1990).

\smallskip

\noindent 7. R.L. Jaffe, {\it Phys. Lett.\/} {\bf B229}, 275 (1989).

\smallskip

\noindent 8. H.-W. Hammer, Ulf-G. Mei{\ss}ner and D. Drechsel,
preprint TK 95 23 (1995).

\smallskip

\end{document}